\documentclass[final]{aipproc}

\layoutstyle{6x9}

\font\msytw=msbm9 scaled\magstep1

\def\xx{{\bf x}}
\def\yy{{\bf y}}
\def\L{\Lambda}
\def\G{\Gamma}
\def\l{\lambda}
\def\s{\sigma}
\def\a{\alpha}
\def\b{\beta}
\def\m{\mu}
\def\th{\theta}
\def\io{\infty}
\def\RRR{\hbox{\msytw R}}
\def\ZZZ{\hbox{\msytw Z}}
\def\NNN{\hbox{\msytw N}}
\def\nn{\nonumber}
\def\sgn{{\rm sgn}}  
\def\media#1{\langle{#1}\rangle}

\def\be{\begin{equation}}
\def\ee{\end{equation}}
\def\bea{\begin{eqnarray}}
\def\eea{\end{eqnarray}}

\begin{document}

\title{Pattern formation in systems with competing interactions\footnote{To 
appear in the AIP conference proceedings of the 10th Granada Seminar on 
Computational Physics, Sept. 15-19, 2008.}}


\author{Alessandro Giuliani}{
  address={Dip.to di Matematica, Universit\`a di Roma Tre, 
L.go S. L. Murialdo 1, 00146 Roma, Italy}}

\author{Joel L. Lebowitz}{
  address={Department of Mathematics and Physics, Rutgers University,
Piscataway, NJ 08854 USA}}

\author{Elliott H. Lieb}{
  address={Departments of Mathematics and Physics,
Princeton University, Princeton, NJ 08544 USA}}

\begin{abstract}
There is a growing interest, inspired by advances in technology, in the low
temperature physics of thin films.  These quasi-2D systems show a
wide range of ordering effects including formation of striped states,
reorientation transitions, bubble formation in strong magnetic fields,
etc. The origins of these phenomena are, in
many cases, traced to competition between short ranged exchange
ferromagnetic interactions, favoring a homogeneous ordered state,
and the long ranged dipole-dipole interaction, which opposes such
ordering on the scale of the whole sample. The present theoretical
understanding of these phenomena is based on a combination of
variational methods and a variety of approximations, e.g., mean-field
and spin-wave theory. The
comparison between the predictions of these approximate methods and
the results of MonteCarlo simulations are often difficult because of
the slow relaxation dynamics associated with the long-range nature of
the dipole-dipole interactions. In this note we will review recent work 
where we prove existence of periodic structures in some lattice and 
continuum model  systems with competing interactions. The continuum 
models have also been used to describe micromagnets, diblock polymers, etc.
\end{abstract}

\maketitle


\section{Introduction}
The formation of mesoscopic (nano/micro) scale patterns (interpreted
broadly) in equilibrium systems is often due to a competition between
interactions favoring different microscopic structures.  As an example suppose 
we begin with a
short ranged potential favoring local ``alignment'' of the microscopic
constituents of the system, e.g. nearest neighbor ferromagnetic
interactions of an Ising spin system.  At low temperature this would
lead to essentially all spins pointing in the same direction.  If we
now add a long range interaction which does not like this ordering
then the system will do the best it can by forming mesoscopic domains
with different aligments.

Such a competition occurs in many systems.  It is well illustrated in
patterns observed in the low
temperature structure of thin films.  These quasi-2D systems show a
wide range of ordering effects including formation of striped states,
reorientation transitions, etc. \cite{DMW00,SW92}, see Figure 1. 

\begin{figure}
  \includegraphics[height=.3\textheight]{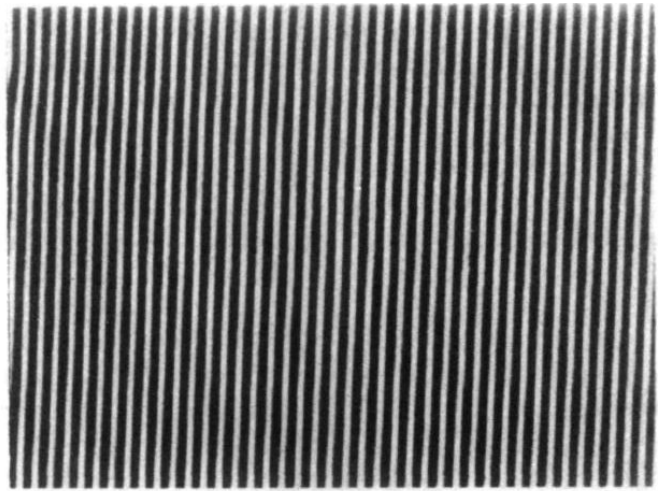}
  \caption{Ferrimagnetic garnet film on GGG at zero magnetic field and
$T=0.6$$T_c$, with $T_c=192^\circ C$. From 
M. Seul and R. Wolfe, PRA (1992)}
\end{figure}

The origins of these phenomena can, in many cases, be traced to the competition
between short ranged exchange (ferromagnetic) interactions, favoring a 
homogeneous ordered state, and long ranged dipole-dipole type
interactions, which oppose such ordering on the scale of the whole
sample \cite{DMW00,SW92,TKNV05,SK06}.

This type of competitive interaction is believed responsible 
for many of the observed patterns in a great variety of systems,
including thin magnetic films \cite{SW92}, micromagnets \cite{Br,DKMO,GD82}, 
diblock copolymers \cite{HS,L80,OK},
anisotropic electron gases \cite{SK04,SK06}, Langmuir monolayers \cite{M88},
lipid monolayers \cite{KM99}, liquid crystals \cite{MS92}, polymer films
\cite{H00}, polyelectrolytes \cite{BE88}, charge-density waves in 
layered transition metals \cite{Mc75}, superconducting films 
\cite{EK93}, colloidal suspensions and cell membranes 
\cite{Destainville1, Destainville2}. Many of these systems are characterized by
low temperature phases displaying periodic
mesoscopic patterns, such as stripes or bubbles. 

The simplest models to describe such systems are Ising or ``soft spin'' models
with a short range ferromagnetic interaction and a power law long range 
antiferromagnetic pair potential
\cite{MWRD95,AWMD95,SS99,LEFK94,GTV00,GD82,CEKNT96}. 
The zero temperature phase diagram of these models has been thoroughly
investigated over the last decade and a sequence of transitions 
from an antiferromagnetic homogeneous state 
to periodic striped or lamellar phases 
with domains of increasing sizes has been predicted, as 
the strength of the ferromagnetic coupling is increased from zero to large
positive values. These theoretical predictions are mostly based on 
a combination of variational techniques and stability analysis: they start by
{\it assuming} a periodic structure, proceed by computing the 
corresponding energy and finally by comparing that energy to the 
energy of other candidate structures, usually by a combination of analytical
and numerical tools. These calculations give an 
excellent account of some of the observed ``universal'' patterns displayed by 
the aforementioned systems. However they run the risk 
of overlooking complex microphases that have not been previously identified 
\cite{BF99}. 
This risk is particularly significant in cases, as those under analysis,
where dynamically (e.g. in Monte Carlo simulations)
the domain walls separating different microphases are 
very long lived as the temperature is lowered \cite{MWRD95}.

To develop a complete ab initio theory of pattern formations it is necessary 
to be able to first {\it prove} 
periodicity of the ground state and then proceed with 
a variational computation within the given ansatz. The problem is not simple. 
Most 
of the mathematically rigorous techniques developed for obtaining the low 
temperature phase diagram of spin systems, e.g., the Pirogov--Sinai theory 
\cite{PS75}, depend on the interaction being short range. Only methods based 
on reflection positivity \cite{FILS78} or on convexity \cite{Hu78,PU78,JM001,
JM002,K99} seem applicable to the kind of potentials considered here.

In this paper we review recent progress on the construction of periodic 
ground states in one and two dimensional spin systems,
both in the case of discrete and continuum systems \cite{GLL1,GLL2,GLL3}. 
The analysis is based on reflection
positivity methods \cite{FILS78}. In the cases where it applies, it gives a 
full justification of the variational calculations based on the 
periodicity assumption.

The paper is organized as follows: in the next section we
describe the class of discrete and continuum spin models we will be concerned 
with; next, we present our main results about the 
zero temperature phase diagram of these models; then we 
give a sketch of the proof and, finally, 
we draw our conclusions.

\section{The models}\label{models}

\subsection{Discrete models}

The simplest discrete spin model describing the class of systems we are 
interested in is an Ising model with the following Hamiltonian:
\be H^{(1)}_\Lambda=-J\sum_{<\xx,\yy>}
\s_\xx\s_\yy+K\sum_{\xx\neq\yy}\frac{\s_\xx\s_\yy}
{|\xx-\yy|^p}\;,\label{2.1}\ee
where: $\L\subset\ZZZ^d$ is a $d$-dimensional cubic box with (say) 
open boundary conditions; the first summation is over all pairs of 
nearest neighbor sites in $\L$, while
the second summation is over pairs of distinct sites of $\L$; if $\xx\in\L$,
$\s_\xx\in\{\pm1\}$ is the Ising spin variable; $p>d$, in order to guarantee
the summability of the long range potential. A physically interesting case
is the one corresponding to $d=2$ and $p=3$, 
in which case the long range term mimics 
the effect of the 3D dipolar interactions among out-of-plane magnetic moments
constrained to a two--dimensional surface as in a thin magnetic film. 

The ground states of (\ref{2.1}) are well understood in two limiting cases: 
if $K=0$, the ground state is ferromagnetic, consisting of spins all 
aligned either up or down; if $J=0$, then the ground state is 
antiferromagnetic, displaying N\'eel order of period 2. 
For general $J$ and $K$ the ground state is not known.
A variational calculation supports the conjecture that in $d=2$, $p=3$, the
ground state is periodic and striped for all $J,K>0$ with $J/K$ larger than
a critical value $j_0$. 
In particular, it is remarkable that 
the checkerboard state has higher energy than the striped one.

Another natural class of discrete spin models is described by the following
Hamiltonian:
\be H^{(2)}_\L = -
J \sum_{\langle \xx, \yy \rangle} \vec{S}_{\xx} \cdot \vec{S}_{\yy}+
K\sum_{\xx \neq \yy} \vec{S}_\xx \hat{W}^{dip} (\xx - \yy) \vec{S}_{\yy} \;,
\label{2.1a}\ee
where $\vec{S}$ are classical Heisenberg spins and 
\be W^{dip}_{ij}(\xx) = -\partial_i \partial_j \frac{1}{|\xx|} = 
\frac{1}{|\xx|^3}\Big(
\delta_{ij} - 3\frac{x_i x_j}{|x|^2} \Big)\;.\label{2.1b}\ee
Also in this case, in the limiting cases $J=0$ or $K=0$ the ground 
state can be determined
exactly: it is ferromagnetic if $K=0$; it displays 
an in-plane uniaxial AF order and is
continuously degenerate if $J=0$. In the general case the ground state is 
unknown.

On the mathematical side, the problem of {\it proving} 
periodicity is very difficult \cite{GLL1,GLL2,GLL3,M,CO}. 
There are very few models where one can
demonstrate rigorously that energy minimizers are periodic.
Note that the problem is not trivial already in 1D.

\subsection{Continuum models}

It is sometimes convenient to consider effective
continuum descriptions of the same spin system.
If $\sigma(x)$ $\in \{\pm 1\}$ and $X_+$ is the region where
$\sigma(x) = +$, one can consider the following model:
\be H^{(3)}_\Lambda = 2J |\partial 
X_+|+K\int_{\Lambda\times\Lambda} d{\bf x}d{\bf y}
\frac {\sigma(x) \sigma(y)} {|x-y|^p+a}\;,\qquad p>d\;.\label{2.2}\ee
The case of ``soft'' local magnetization $\phi(x) \in \RRR$ 
is often considered, too:
\be H^{(4)}_\Lambda = \int_{\Lambda} dx \Big[|\nabla \phi|^2 + F
  (\phi(x))\Big]+ K \int_{\Lambda \times \Lambda} d {\bf x}d{\bf
  y} \frac {\phi(x)\phi(y)} {|x - y|^p + a}\;.\label{2.3}\ee
In the two equations above $a>0$ is an ultraviolet regulator, 
playing the role of the lattice spacing in the previous discrete model. 
In the case of the soft spin model, $F$ is a symmetric 
double well potential, with two minima located at $\phi=\pm\phi_0\neq0$:
typical examples to keep in mind are 
$F(\phi)=(\phi^2-\phi_0^2)^{2}$ or $F(\phi)=(|\phi|-|\phi_0|)^{2}$ or
\be F(\phi)=\left\{\matrix{
-\frac{\phi^2}2-\frac1\b I(\phi)\,, \hskip1.truecm {\rm if}\ |\phi|<1
\,,\cr +\io\;,\hfill 
\hskip1.2truecm {\rm if}\ |\phi|\ge 1\;, \cr}\right.\qquad 
\label{2.4}\ee
where $I(\phi)=-\frac{1-\phi}2\log\frac{1-\phi}2-\frac{1+\phi}2\log
\frac{1+\phi}2$.
The gradient term in (\ref{2.3}) represents the cost of a
transition between two phases (the homogeneous magnetized phases 
induced by the short range exchange interaction), 
while the term $F$ represents the local
free energy density of a homogeneous system in a mean field approximation.

Similarly to the discrete case, the minimizers of (\ref{2.2}) and (\ref{2.3})
can be easily determined in the limiting cases where only the attractive
or only the repulsive interactions are present. In the presence of a 
competition, it is again conjectured, on the basis of variational computations,
that the minimizers are periodic and that they display periodic striped
(or lamellar) order \cite{SK04}.

Both the models in this and in the previous section can be extended to the case
of non zero magnetic field, in which case the expected phase diagram is even 
more complex, e.g., periodic bubbled states are expected at high enough 
magnetic field \cite{GD82}. The rigorous results concerning these more 
complicated cases are still very partial, and we will not 
consider them explictly in this paper.

\section{Main results}\label{results}

We have proven a number of results in the 1D case, both for the discrete and 
the continuum models. The higher dimensional case is in many respects still 
open. Partial results include an example of a 2D dipole system with in-plane 
dipoles, reminiscent of model (\ref{2.1a}), that, in the presence
of a special nearest neighbor ferromagentic interaction, has periodic 
ground states displaying striped periodic order. In the following we want 
to describe these results, starting from the case of one dimension. 

\subsection{One dimension}

In one spatial dimension, we have a quite complete picture of the ground 
states of models $H^{(1)}_\L$, $H^{(3)}_\L$ and $H^{(4)}_\L$. Regarding 
the models with Hamiltonian $H^{(1)}_\L$ and $H^{(3)}_\L$, our main result 
can be summarized as follows \cite{GLL1}.\\
\\
{\bf Theorem 1.} {\it Let $d=1$. 
For any $J, K \ge 0$ and any $p>1$, the specific ground state 
energies $E_0^{(1)}(\L)$ and $E^{(3)}(\L)$ of, respectively, 
$H_\Lambda^{(1)}$ and $H^{(3)}_\L$ in the thermodynamic limit are given by:
\bea &&
\lim_{|\Lambda| \rightarrow \infty} \frac {1} {|\Lambda|} E_0^{(1)}(\Lambda) =
\inf_{h\in\NNN}e_1(h)\;,\nn\\
&& \lim_{|\Lambda| \rightarrow \infty} \frac {1} {|\Lambda|} E_0^{(3)}
(\Lambda) = \inf_{h\in\RRR_+}e_3(h)\;,
\eea
where $e_1(h)$ and $e_3(h)$ are the specific energies of a striped periodic 
configuration
of period $2h$, obtained by extending periodically over the whole line 
the functions $\s_x=\sgn(x)$, $x=-h+1,\ldots,h$, and $\s(x)=\sgn(x)$, 
$x\in(-h,h]$, respectively. In the presence of periodic boundary conditions,
the only ground states of $H^{(3)}_\L$ are the optimal periodic
striped configurations; in the case of the discrete Ising model, the same
conclusion is valid, if the ring has a length divisible by the optimal
period.}\\

{\bf Remarks.}\\
1) The optimal period $2h$ increases with $J$; when $J=0$, $h=1$ in the 
discrete and $h=0$ in the continuum, i.e., as $J\to0$ the oscillations
become wild and the period shrinks to zero.  
Depending on the
values of $p$, $J$ and $K$, the ground state has finite periodicity or 
is ferromagnetic:
if $p >2$ and $J/K$ smaller than a suitable $j_c$ the period is finite
and diverges at $j_{c}$; if $J/K \ge j_c$ the ground state is FM.
If $1 < p \leq 2$, then $j_c = \infty$ 
and the period will increase as $J/K$ increases, becoming 
{\it mesoscopic/macroscopic}.
\\
2) The proofs are based on a generalized notion of reflection positivity
(RP), see \cite{FILS78} and below for the description 
of reflection positivity. In the present case one
needs first to describe the states as collections of blocks, and
then apply RP to the effective model of interacting blocks.\\
3) The proof works equally well for a larger class of long range potentials
$v(r)$: it is enough that $v(r)$ is the Laplace transform of a positive
measure, i.e., $v(r)=\int_0^\io d\a\,\m(\a) e^{-\a|r|}$. 
This class of interactions
include, besides the power laws, the exponentials and arbitrary positive 
linear combinations of exponentials.\\
4) Inclusion of temperature is not trivial: block RP is
lost for $\beta<+\infty$. 
One expects for $p>1$ a \underline{unique} Gibbs measure,
whose typical configurations are close to the periodic
ground states determined above.  This is ``easy'' to prove for $J=0$ and 
$p>2$ and is 
probably not true for $p<1$ when the potential is not absolutely summable.
(\underline{Note}: for purely {\it ferromagnetic} power law potentials
$v(r)\sim-1/r^p$, there is a finite temperature spontaneous magnetization 
for $1<p\leq2$).\\
5) As mentioned above, we are unable at present to extend our results 
to the case where there is a magnetic field acting on the system or
 there is a non zero specified magnetization.  
This is true both for the lattice, where very complex structures are expected
(e.g., for $J=0$ the existence of a ``devil's staircase'' has been proved
\cite{PerBak}), and for the continuum where one expects 
simple periodic structures with blocks of alternate lengths \cite{Huse08}.\\

The proof of Theorem 1 works exactly in the same way both for $H_\L^{(1)}$
and for $H_\L^{(3)}$. The extension to $H_\L^{(4)}$
requires a refinement of the block RP ideas, see 
\cite{GLL3}. The final result, stated in an informal way, is the following 
(see \cite{GLL3} for a mathematically rigorous statement, making precise the 
choice of boundary conditions and the notion of infinite volume minimizers).
\\
\\
{\bf Theorem 2.} {\it If $d=1$, all the minimizers of $H_\L^{(4)}$ 
are either {\it simply 
periodic}, of finite period $T$, 
with zero average, or of constant sign (and are constant if $F$ is convex 
on $\RRR^+$). By ``simply periodic'' we mean that within a period the minimizer
has only one positive and one negative region, with the negative part obtained
by a reflection from the positive part.}
\\

{\bf Remarks.}\\
1) Similarly to the ``hard'' spin case, the proof works even if the power law 
potential is replaced by a $v(r)$ that is the Laplace transform of a positive 
measure, in particular in the case of an exponential interaction. 
Of course, as in Theorem 1, the value of the period and, 
in this case, the shape
of the minimizer within one period, is obtained by a variational computation,
whose result depends on $v(r)$, $J$ and $K$.\\
2) One can provide explicit examples of cases where the system displays 
a transition from a homogeneous to a periodic non homogeneous state. For
instance, if $F(\phi) = (|\phi| - 1)^2$ and $v(x)=\lambda e
^{-|x|}$, the previous result combined with an explicit computation
imply that the minimizer is constant for $\lambda \leq 3/2$ and
periodic with zero average for $\lambda > 3/2$.\\
3) The result of Theorem 2 can be extended to a larger class of 
free energy functionals. In particular, it can be applied to
the study of effective 1D models for martensitic phase transitions, see 
\cite{KM,M}.

\subsection{Two dimensions}

As mentioned above, the case of two or more dimensions is in many respects 
still open. Partial results include:\\
1) lower bounds on the specific ground state energy of (\ref{2.1})
and (\ref{2.2}), agreeing at lowest order with the energy of the striped case;
\\
2) an example of a 2D dipole system with in-plane dipoles
and a special exchange interaction, whose ground states can
be proved to be striped and periodic.\\
\\
For the precise statement and the proof of claim (1), we refer to \cite{GLL1}
(let us just mention that the proof is based on apriori bounds on the energy 
of Peierls contours). Here we want to describe in more detail the result 
mentioned in item (2), which is, as far as we know, the only example of a 
spin model in two or more dimensions with real dipole interactions, 
for which existence of periodic striped order has been proved.

The two dimensional spin model that we consider is a modification of model
$H^{(2)}_\L$, defined by the following Hamiltonian:
\be \widetilde H^{(2)}_\Lambda 
= \sum_{\xx \neq \yy} \vec{S}_\xx \hat{W}^{dip} (\xx-\yy) 
\vec{S}_\yy - \sum_{\langle \xx,\yy \rangle} \Big( 
J \vec{S}_x \cdot \vec{S}_y +
\lambda (\vec{S}_x \cdot \vec{S}_y)^2 \Big)\;,\label{2D}\ee
where $\L$ is a 2D square lattice, $\lambda \geq 0$ and 
$\vec{S}_\xx$ are in-plane spins, whose allowed
directions are only $\uparrow$, $\downarrow$, $\rightarrow$ and $\leftarrow$.
Note that the $\lambda$ term has the effect of encouraging alignment or 
antialignment but
this term alone cannot create periodic order. Our main result can be summarized
as follows.
\\
\\
{\bf Theorem 3.} {\it Let $d=2$. 
For $J \geq 0$ and $\lambda$ large enough, the specific
ground state energy of $\widetilde H^{(2)}_\lambda$ 
in the thermodynamic limit is given by:
\be \lim_{|\Lambda| \rightarrow \infty} \frac {1}{|\Lambda|} 
\widetilde E^{(2)}_0 (\Lambda)= \min_{h \in \ZZZ+} \tilde e_2(h)
\;,\ee
where $\tilde e_2(h)$ is the specific energy of a striped configuration of
period $2h$, consisting of stripes of uniformly polarized columns,
all of size $h$, and with alternate up/down polarization.
On a torus of side divisible by the optimal period, the
only ground states are the optimal periodic striped configurations, either 
displaced vertically or horizontally.}
\\
\\
{\bf Remarks.}\\
1) The condition on $\lambda$ is not uniform in $J$.  
It is unclear whether the same result should be valid for large $\lambda$,
uniformly in $J$, or even up to $\lambda = 0$.\\
2) The proof of Theorem 3 is based on an extension of the ideas of the proof
of Theorem 1. We first show that, for $\l=+\io$, the system preferes to 
have the columns all completely polarized (because of the anisotropy
of the dipolar potential); this means that, for the purpose of computing 
the ground state energy, we can restrict to 1D confirgurations of up or down 
columns, and at this point we can apply Theorem 1. Next we show that if
$\l$ is sufficiently large, then in the ground state there are no 
perpendicular neighboring spins, and this concludes the proof. See \cite{GLL2}
for details.

\section{Reflection positivity}\label{proof}

As mentioned in previous sections, the proofs of Theorem 2 and 3 extend the 
ideas of the proof of Theorem 1, which is based on a generalized notion of 
reflection positivity. Let us clarify here what we mean by reflection 
positivity, and how can one apply it to the problem of determining the
ground state of $H^{(1)}_\L$, at least in the simple case $J=0$.

Let us consider the Hamiltonian 
\be H_\L=\sum_{-N< x<y\le N}\s_x\s_y\, v(y-x)\;,\ee
in the presence of periodic boundary conditions and with $v(x)=\sum_{n\in\ZZZ}
|x+2nN|^{-p}$. Note that, if $x>0$, using that $x^{-p}=\Gamma(p)^{-1}
\int_0^\io d\a \a^{p-1} e^{-\a x}$, we can rewrite:
\be v(x)=\int_0^\io \frac{d\a}{\G(p)} \frac{\a^{p-1}}{1-e^{-2\a N}}(e^{-\a x}
+e^{-\a(2N-x)})\;.\ee
Therefore,
\be H_\L(\underline\s)=H_L(\underline\s)+H_R(\underline\s)
+\sum_{x=-N+1}^0\sum_{y=1}^{N}
\int_0^\io d\a\, \m(\a)\s_x\s_y\, 
(e^{-\a(y-x)}+e^{-\a(2N-y+x)})
\;,\label{3.1}\ee
where: $\m(\a)=\frac{1}{\G(p)} \frac{\a^{p-1}}{1-e^{-2\a N}}$,  
$\underline\s=(\s_{-N+1},\ldots,\s_{-1},\s_0,\ldots,\s_N)$ and
\bea && H_L(\underline\s)=\sum_{-N<x<x'<0} \s_x\s_{x'}\, v(x'-x)\;,\nn\\
&& H_R(\underline\s)=\sum_{0\le y<y'\le N} \s_y\s_{y'}
\, v(y'-y)\;.\eea
If $\th\underline\s$ is the configuration with 
$(\th\underline\s)_i=-\s_{-i+1}$ and $A_\a(\underline\s)$ is defined as
\be A_\a(\underline\s)=\sum_{y=0}^N\s_y e^{-\a y}\;,\ee
we can rewrite (\ref{3.1}) in the form:
\be H_\L(\underline\s)=H_L(\underline\s)+H_R(\underline\s)
-\int_0^\io d\a \m(\a) e^\a \big[A_\a(\underline\s) A_\a(\th\underline\s) 
+e^{-2\a N} A_{-\a}(\underline\s) A_{-\a}(\th\underline\s)\big]
\;,\label{3.2}\ee
Now the crucial remark is that the integral in the right hand side of 
(\ref{3.2}) defines a scalar product between the configurations $\underline\s$
and $\th\underline\s$. Therefore, defining
\be \media{\underline\s_1,\underline\s_2}=\int_0^\io d\a\,\m(\a) 
e^\a \big[A_\a(\underline\s_1) A_\a(\underline\s_2) 
+e^{-2\a N} A_{-\a}(\underline\s_1) A_{-\a}(\underline\s_2)\big]\ee
and using that, for any scalar product,
\be \media{\underline\s_1,\underline\s_2}\le \frac12\Big(
\media{\underline\s_1,\underline\s_1}+\media{\underline\s_2,\underline\s_2}
\Big)\;,\ee
we get 
\be H_\L(\underline\s)\ge \frac12\Big(H_L(\underline\s)-\media{\th\underline\s,
\th\underline\s}+H_R(\th\underline\s)
\Big)+\frac12\Big(H_L(\th\underline\s)-\media{\underline\s,\underline\s}
+H_R(\underline\s)\Big)\;,\ee
which means 
\be H_\L(\underline\s)\ge\frac12 \big(H_\L(\underline\s_L)+H_\L(\underline\s_R)
\big)\;,\ee
with $\underline\s_L=(\s_{-N+1},\ldots,\s_0,-\s_0,\ldots,\-\s_{-N+1})$ and
$\underline\s_R=(-\s_{N},\ldots,-\s_1,\s_1,\ldots,\s_{N})$. In other words, 
given any configuration $\underline\s$, at least one of the two configurations 
obtained from $\underline\s$ by reflection around $(0,1)$ has better (or equal)
energy than the original configuration. So, if we want to reduce the energy, we
can keep reflecting about all possible bonds; proceeding like this we end up 
with the configuration $\cdots+-+-+-\cdots$

In the presence of a nearest neighbor ferromagnetic interaction, we proceed
in a similar fation, but we only reflect about the bonds separating a plus 
from a minus spin. After repeated refelctions we are left with a configuration 
consisting of a sequence of blocks of polarized spins, all of the same size 
and with alternate polarization. For more details, see \cite{GLL1}
(see also \cite{GLL2} and \cite{GLL3} for a corrected discussion about 
the checkerboard estimate).

\section{Conclusions}\label{conclusions}

In this note we reviewed some recent rigorous results about existence of
periodic striped states for a number of 1D and 2D spin systems, described
by discrete or continuum models. The proofs are based on a generalized notion 
of reflection positivity, and require the long range interaction to be 
{\it reflection positive}, i.e., the Laplace transform of a positive measure.
The short range interaction needs to be among nearest neighbor sites. The proof
applies to 1D systems or higher dimensional systems which can be proven
to display 1D ground states by apriori energy estimates. Open problems
include the inclusion of magnetic fields, of a non zero temperature and,
most importantly, the proof that the ground states of 
the $d$--dimensional spin systems 
described by Eqs. (\ref{2.1}), (\ref{2.1a}), (\ref{2.2}) and (\ref{2.3})
are translational invariant in $d-1$ coordinate directions. 
So far, this claim has only been proven for $\widetilde H_\L^{(2)}$ in 
(\ref{2D}). There are 
good hopes to extend the proof to a new class of 2D models, relevant for 
the description of martensitic phase transitions \cite{KM}. 

Let us also mention that these models with competing interactions 
are closely related to a class of 
systems considered by Lebowitz and Penrose in \cite{LP}.  They
considered the case when the pair potential is the sum of a short
range interaction, $\phi(r)$,  favoring phase segregation on the 
macroscopic scale and a long range (Kac type) interaction favoring a
uniform density, e.g. $v_\gamma (r) = \alpha \gamma^d \exp\{-\gamma r\},
\alpha>0$, $d$ the space dimension.  Lebowitz and Penrose proved that,
in the limit $\gamma \rightarrow 0$, 
this competition will result in the system breaking up
into a ``foam'' consisting of mesoscopic regions of the
different phases.  These will  have a characteristic length large compared to
that of the short range potential and small compared to that of
the long range potential $\gamma^{-1}$.  This may give rise to the
different kinds of patterns 
observed experimentally in many systems.  We have shown in 
\cite{GLL2} that in one dimension when the long range Kac potential is of 
the exponential type,
or more generally is reflection-positive, then these droplets will
form a periodic ground states with period of order $\gamma^{-2/3}$.  A
heuristic argument suggests that the scaling of the patterns in three
dimensions will  be like $\gamma^{-4/5}$ \cite{SF03}.

As pointed out in \cite{SF03} the effect of the long range repulsive
interaction is similar to that of surfactants which lower the surface
tension between the two phases.  We plan to explore further the general
phenomena of pattern formation
due to competing interactions which occurs in many systems
beyond those described earlier, e.g. Langmuir monolayers, lipid monolayers,
liquid crystals, two dimensional electron gases, diblock copolymers,
etc. This type of competition may also be responsible for some of the
phenomona observed in confined water \cite{38} and in aqueous surfactant
solutions \cite{39}. A related kind of competition, due to geometric 
frustration, can induce the formation of striped periodic patterns 
in buckled collodial monolayers \cite{40}.  
As pointed out in \cite{40,SL08}, this is related to 
stripe formation in compressible antiferromagnents on a triangular lattice 
\cite{41}.

\begin{theacknowledgments}
The work of JLL was supported by NSF Grant DMR-044-2066 and by AFOSR
Grant AF-FA 9550-04-4-22910. The work of AG and EHL 
was partially supported by U.S. National Science Foundation
grant PHY-0652854. JLL would like to thank Pedro Garrido and Joaquin Marro
for their hospitality at their conference in Granada.
\end{theacknowledgments}

\end{document}